\documentclass[twocolumn]{autart}    % Enable this line and disable the  autart
\usepackage{graphicx}          % Include this line if your 
\usepackage{amsmath}
\usepackage{array}
\usepackage{comment}
\usepackage{graphicx}
\usepackage{caption}
\usepackage{subfigure}
\usepackage{cases}
\usepackage{amssymb}
\usepackage{enumerate}
\usepackage{cite}
\usepackage{natbib}
\usepackage{wrapfig}
\usepackage{graphicx}
\setcitestyle{authoryear,open={(},close={)}}

\edef\endfrontmatter{%
  \unexpanded\expandafter{\endfrontmatter}% the current code
  \noexpand\endNoHyper % add \endNoHyper at the end to match \NoHyper
}
\usepackage[colorlinks,
            linkcolor=blue,
            anchorcolor=blue,
            citecolor=blue
            ]{hyperref}
\usepackage{color}
\definecolor{r}{rgb}{0.86, 0.08, 0.24}

\begin{document}

\begin{frontmatter}

\title{Dynamics of coupled implicit and explicit opinions\thanksref{footnoteinfo}} % Title, preferably not more 

\thanks[footnoteinfo]{This paper was not presented at any IFAC 
meeting. Corresponding author: Xiaofan Wang.}

\author[Paestum]{Qi Zhang}\ead{douuu@sjtu.edu.cn},    % Add the 
\author[Paestum]{Lin Wang}\ead{wanglin@sjtu.edu.cn},               % e-mail address 
\author[Paestum,Baiae]{Xiaofan Wang}\ead{xfwang@sjtu.edu.cn}  % (ead) as shown

\address[Paestum]{Department of Automation, Shanghai Jiao Tong University, and Key Laboratory of System Control and Information Processing, Ministry of Education of China, Shanghai 200240, China}  % Please supply                                              
%\address[Rome]{Senate House, Rome}             % full addresses
\address[Baiae]{Department of Automation, School of Mechatronic Engineering and Automation, Shanghai University, Shanghai 200444, China}        % here.

                             % keyword list or with the 
                                          % help of the Automatica 
                                          % keyword wizard

\begin{abstract}                          % Abstract of not more than 200 words.

This paper proposes a dual opinions co-evolution model based on the dual attitudes theory in social psychology, where every individual has dual opinions of an object: implicit and explicit opinions. The implicit opinions are individuals' inner evaluations affected by the experience, while the explicit opinions are external expressions of the evaluations and are influenced by public opinion. We theoretically analyze the dynamics of coupled dual opinions in an adversarial discussion scenario, in which two opposing groups participate in the discussion. We rigorously give a method to grade the degree of individual bias. 
When individuals have a slight bias, dual opinions reach a consensus, showing the phenomenon of acceptance. When individuals are moderately or even extremely biased, they will exhibit the phenomenon of outward compliance, that is, their publicly explicit opinions conform to public opinion, while implicit opinions will reach a persistent disagreement and even polarization. Our investigation indicates that dual opinions may not be monotone initially but eventually converge monotonically. Besides, we analyze the influence of parameters on the co-evolutionary process and results. Numerical simulations of the dual opinions formation process produce inward and outward conformity phenomena, namely acceptance and compliance.
\end{abstract}

\end{frontmatter}

\section{Introduction}
Sensitive topics such as same-sex marriage \citep{flores2016backlash}, global warming \citep{mccright2011politicization}, and abortion \citep{mouw2001culture} often spark heated debates. This naturally brings up a question: how group discussion influences our opinions? Different empirical studies arrive at different answers. Some studies show that group discussion intensifies members' initial inclination and leads to polarization \citep{myers1970discussion}.
Other studies suggest that communication could narrow differences. For example, positive interactions could reduce racial prejudice \citep{greeley1971attitudes,sigelman1993contact}. 
However, some evidence indicates that such changes usually only occur in explicit attitudes that are susceptible to the environment \citep{wilson2000model}.
Inspired by these phenomena, we wish to investigate the co-evolution of implicit and explicit opinions in group discussion from the perspective of opinion dynamics.

The concepts of implicit and explicit attitudes come from the dual attitudes theory in social psychology proposed by \cite{wilson2000model}, which suggests that individuals have two different evaluations of an object simultaneously. 
{Implicit attitudes are affected by individuals' experiences}.
When individuals realize that their implicit attitudes may be illegitimate or unwanted, they will consciously make explicit attitudes more in line with social expectations to override their implicit attitudes.
\cite{wilson2000model,fiske1998stereotyping} indicate that stereotyping and bias exist at an implicit and automatic level.
Accordingly, in some cases, the implicit bias of individuals and the social environment jointly affect the co-evolution of dual attitudes. On the other hand, although the terms ``attitudes" and ``opinions" have different definitions in sociology, we can roughly understand these terms as the degree of positive or negative cognitive orientation to an object \citep{eagly1993psychology}. Therefore, the current research on opinion dynamics does not distinguish them and usually calls them ``opinions" \citep{friedkin2015problem,proskurnikov2017tutorial}, and so does this paper.

Opinion evolution has attracted much attention from researchers, and many remarkable dynamics models have been proposed to explain the mechanism of opinion change in discussion, such as the DeGroot model \citep{proskurnikov2015opinion}, the Friedkin-Johnsen (F-J) model \citep{ye2019influence},
the Bounded Confidence (BC) model \citep{luo2022Adapted,cheng2019opinion}, 
and the biased assimilation model \citep{dandekar2013biased,xia2020analysis}.

\textit{Co-evolutionary models}: The rise of co-evolution research in recent years stems from the proposal of the EPO model framework in \cite{ye2019influence}. In this framework, the private and expressed opinions co-exist on an object.
The private opinion is unobservable for others in the co-evolution. The expressed opinion is the weighted average of the private opinion and public opinion, and the weight is affected by group pressure.
\cite{ye2019influence} use the F-J model as the private opinion update model describing the influence of initial bias and 
stubbornness on the opinion evolution. 
Following this, \cite{cheng2019opinion,hou2021opinion,cheng2021social,luo2022Adapted} modify the BC model to study the co-evolution of opinions.
\cite{hou2021opinion} show that the number of final opinion clusters is inversely proportional to the growth rate of the stubbornness of initial bias.
\cite{cheng2019opinion} indicate that social pressure can reduce  the disagreement of the group. 
\cite{cheng2021social,luo2022Adapted} show that an appropriate pressure level leads to the group consensus, and as the group pressure increases, the group will split  into clusters.
\cite{cheng2021social} {reproduce two types of conformity phenomena called acceptance and compliance, where acceptance is that all individuals’ expressed opinions and private opinions reach a consensus, and compliance means that individuals' expressed opinions are more in line with public opinion than private opinions.}

Using the F-J model \citep{ye2019influence} or the BC model \citep{cheng2019opinion,cheng2021social,hou2021opinion,luo2022Adapted}  as the update rule of private opinions fails to capture opinion polarization, i.e., opinions converge to the extreme value of the state space, since the weighted average update rule decreases the gap between the biggest and smallest opinion values. Motivated by this, this paper will use the biased assimilation model \citep{dandekar2013biased} as the implicit opinion dynamic model, which can generate classic patterns of opinion evolutionary results, including consensus, disagreement, and polarization, and shows the phenomena of inward and outward conformity, namely acceptance and compliance. 

This paper proposes a co-evolutionary model of dual opinions based on the EPO model framework \citep{ye2019influence}, where each individual has implicit and explicit opinions of an object simultaneously. The evolution of implicit opinions follows the biased assimilation model \citep{dandekar2013biased}, in which individuals process new information in a biased way to support their evaluation. The explicit opinions are consciously activated to conform to public opinion and are tightly coupled with implicit opinions. On the other hand, implicit opinions are private and unobservable, so individuals can only receive explicit opinions from their neighbors.
We theoretically analyze the dual opinions dynamic model on the two-island networks in an adversarial discussion scenario, where two opposing groups participate in the discussion,  such as political debates between Democrats and Republicans. 

We conduct a theoretical analysis of the dual opinions co-evolutionary process and results under different levels of individual bias. We rigorously give a method to distinguish between three degrees of individual bias, i.e., slight, moderate, and extreme bias. 
When individuals are slightly biased, the dual opinions reach a consensus at the neutral state, showing a phenomenon of acceptance as in \cite{myers_social_2010}, that is, individuals sincerely believe and express in line with the public opinion. When individuals have moderate or extreme bias, the implicit opinions of individuals belonging to opposing groups converge but do not reach a consensus. Due to social pressure, individuals express explicit opinions that are more in line with public opinion than implicit opinions, showing a phenomenon of compliance as in \cite{myers_social_2010}. If individuals are {extremely biased}, the group discussion will deepen the gap of implicit opinions between opposing groups, resulting in opinion polarization.
Our investigations also show that explicit opinions are always more neutral than implicit ones and will never converge to the extreme states in  evolution. Besides, we analyze the impact of parameters such as the biased parameter, group pressure of the social network, and initial explicit opinions on the co-evolutionary process and result {in the adversarial discussion scenario}. Numerical simulations show the process of dual opinions co-evolution.

The outline of the paper is as follows. First, Section \ref{sec:model} introduces the dual opinions evolution model. Then,  Sections \ref{sec:theoretical} and \ref{sec:numerical} present theoretical and numerical results respectively. Finally, Section \ref{sec:conclusion} concludes this paper.
Detailed proofs and lemmas are in the Appendix.

\textbf{Notations}: $W=\{w_{ij}\}_{n\times n}\in \mathbb{R}^{n\times n}$ denotes the matrix $W$ with elements $w_{ij}\in \mathbb{R}$. $|S|$ is the cardinality of the set $S$. 
$\mathbb{Z}_+$ is the set of positive integers.
$\mathcal{P}_l=(p_{l,1},\cdots, p_{l,k})$ is a tuple with $l\in \mathbb{Z}_+$. {We say} $\mathcal{P}_{l_1}>\mathcal{P}_{l_2}$ if  {$\mathcal{P}_{l_1}$ and $\mathcal{P}_{l_2}$ have the same cardinality $k$ and $p_{l_1,i}> p_{l_2,i}$ holds for all $i\in\{1,\cdots,k\}$}.

\section{The dual opinions co-evolution model}\label{sec:model}

Before introducing the dual opinions co-evolution model, we introduce some concepts in graph theory.

\textbf{Graphs}:
$G(W)=(V,E,W)$ represents the communication network, where $V=\{1,2,\cdots,n\}$ is the set of agents, and $E=\{e_{ij}:i,j \in V\}$ is the edge set. $W=\{w_{ij}\}_{n\times n}$ is the adjacency matrix of the graph $G(W)$. We assume that $G(W)$ is an undirected network, i.e., if $e_{ij}\in E$, then $e_{ji}\in E$ and $w_{ij}=w_{ji}=1$, otherwise,  $w_{ij}=w_{ji}=0$. $N_i=\{j\in V: e_{ij}\in E, j\neq i\}$ is the neighbor set of individual $i\in V$.

Inspired by the dual attitudes theory \citep{wilson2000model}, this paper considers that every individual $i\in V$ has dual opinions: implicit and explicit opinions, denoted as $x_i(t)\in [0, 1]$ and $y_i(t)\in [0, 1]$, respectively. 
The dual opinions $x_i(t), y_i(t)$ of individual $i\in V$ can be interpreted as the degree of support for a statement.
Extreme opinion values of $0$ and $1$ indicate disagreement and support for the statement, respectively, and $0.5$ indicates the neutral opinion.
The implicit opinion is activated autonomously and influenced by experiences. \cite{fiske1998stereotyping,wilson2000model} indicate that prejudice and stereotyping exist at an implicit and automatic level, revealing that implicit bias could influence the evolution of implicit opinions in some cases. Therefore, we use the biased assimilation model \citep{dandekar2013biased} to update implicit opinions, in which implicit opinions evolve in a biased way to maintain the prior evaluation.  When individuals realize their implicit opinions are untimely or unwanted, they will consciously motivate an explicit opinion more in line with social expectations to override their implicit opinions \citep{wilson2000model}.
In other words, explicit opinions are governed by implicit and public opinions. The dual opinions update rules are as follows.

At each time step $t>0$, the individual $i\in V$ updates its implicit and explicit opinions according to
\vspace{-0.5cm}
\begin{small}
\begin{equation}\label{eq:sys}
	\begin{split}
		&x_i(t+1)= \frac{x_i(t)^{b_i}s_i(t)}{x_i(t)^{b_i}s_i(t)+(1-x_i(t))^{b_i}(d_i-s_i(t))}~,\\
		& y_i(t+1)=\phi_ix_i(t+1)+(1-\phi_i)\hat{y}_{i,avg}(t),\\
		& \hat{y}_{i,avg}(t)=\sum_{j\in N_i}m_{ij}y_j(t),
	\end{split}
\end{equation}
\end{small}
where $b_i>0$ is the bias parameter, $s_i(t)=\sum_{j\in N_i}w_{ij}y_j(t)$ is the sum of explicit opinions from neighbors received by individual $i\in V$, $d_i=\sum_{j\in N_i}w_{ij}$ is the total influence of $i$'s neighbors, and $\phi_i\in(0,1)$ is the resilience to resist social pressure of conforming with public opinion (or resilience for short), as in \cite{ye2019influence}. Likewise, $1-\phi_i$ refers to the group pressure of the social network  (or {social} pressure for short) for individual $i\in V$.  $\hat{y}_{i,avg}(t)$ is the public opinion viewed by individual $i\in V$ at time $t$. 
The matrix $M=\{m_{ij}\}_{n\times n}$ is row stochastic, i.e., $\sum_{j\in N_i}m_{ij}=1$ for $i\in V$.
We assume that for $ j\in V, j\neq i$,
$m_{ij}>0 \Leftrightarrow  w_{ij}>0$.
Consequently, the model \eqref{eq:sys} is now well-defined.
The more detailed explanations of the biased assimilation model refer to the earliest paper \citep{dandekar2013biased}. The update rules of implicit and explicit opinions are based on the EPO model framework \citep{ye2019influence}.

\begin{rem}
According to the opinion update rules \eqref{eq:sys}, implicit and explicit opinions are tightly coupled, which can not be easily decoupled. Unlike the classic biased assimilation model \citep{dandekar2013biased}, in this paper,
individuals can only receive explicit opinions from their neighbors since individuals' implicit opinions are unobservable. Thus, the opinions exchanged between individuals are explicit opinions. 
If $y_i(0)=x_i(0)$ and $\phi_i=1$ hold for all individuals $i\in V$, then their explicit and implicit opinions are the same, i.e., $x_i(t)=y_i(t)$ holds for any $t\geq 0$, and the model \eqref{eq:sys}  degenerates into the classic bias assimilation model \citep{dandekar2013biased}.
\end{rem}

\begin{rem}
The settings of matrix $M$ are varied. \cite{cheng2019opinion,cheng2021social,luo2022Adapted} assume that all individuals' external opinions are observable and thus define that  $m_{ij}=\frac{1}{n}$ for any $i,j\in V$. 
In large-scale networks, it is difficult for individuals to obtain information about everyone, so we assume that $m_{ij}=w_{ij}/d_i$ for any $i,j \in V$, which means that the public opinion viewed by individual $i$ is governed by its neighbors' explicit opinions. 
This assumption is also mentioned in \cite{ye2019influence}.
\end{rem}

Similar to \cite{dandekar2013biased}, the implicit opinion formation process is polarizing if the network disagreement index $\eta(t+1)\geq \eta(t)$ for {all} $t\geq0$, where $\eta(t)=\sum_{(i,j)\in W}w_{ij}(x_i(t)-x_j(t))^2$. Furthermore, we define opinion polarization as a property of evolutionary outcomes, indicating that opinions converge to extreme states $0$ or $1$.

\section{Analyses of the co-evolution of dual opinions}\label{sec:theoretical}

This section will investigate the co-evolution of dual opinions in a kind of adversarial discussion scenario, where two opposing groups of individuals  participate in the discussion, and the individuals in the different groups have opposite initial opinions. The group discussions in this kind of adversarial scenario happen all the time, ranging from whether a vegetarian diet is good for the environment to supporting Democrats or Republicans in US presidential elections.

The research \citep{McPherson2001birds} indicates that similarity brings connection, which implies that more interactions are within the groups. Therefore, we use the two-island network \citep{dandekar2013biased} to model the communication network of two opposing groups in the ``adversarial" environment since the individuals in the same group have similar opinions. 
\begin{defn}
\citep{dandekar2013biased} Given integers $n_1,n_2\geq 0$ and real numbers $p_s,p_d\in(0,1)$, a $(n_1,n_2,p_s,p_d)$-two-island network is a weighted undirected graph $G=(V_1,V_2,E,W)$, where $n_1p_s,n_1p_d,n_2p_s$, $n_2p_d\in \mathbb{Z}_+$, and
\begin{itemize}
    \item $|V_1|=n_1$, $|V_2|=n_2$, and $V_1\cap V_2=\emptyset$.
    \item Each node $i\in V_1$ has $n_1p_s$ neighbors in $V_1$ and $n_1p_d$ neighbors in $V_2$.
    \item Each node $i\in V_2$ has $n_2p_s$ neighbors in $V_2$ and $n_2p_d$ neighbors in $V_1$.
    \item $h_G=p_s/p_d$, $p_s>p_d$, is the degree of homophily of $G$.
\end{itemize}
\end{defn}
In the two-island network, intra-group connections are more than inter-group connections.
Fig.\ref{fig:y0}(a) shows a $(6,6,2/6,1/6)$-two-island network.
The degree of homophily $h_G$ describes the ratio of in-group neighbors to out-group neighbors. The larger $h_G$ is, the greater the influence of in-group neighbors on the evolution of dual opinions. Social pressure $1-\phi_i$ describes the influence of public opinion on the explicit opinion of individual $i\in V$, where the influence originates from {its neighbors related to the social network}.
The coupled nature of the implicit and explicit opinions and the strong non-linearity 
make the co-evolutionary process complex. Therefore, we simplify the complexity of the theoretical analysis by making the following assumption.

\begin{assum}\label{ass} 
	The communication network $G(W)=(V,E,W)$ is a $(n,n,p_s,p_d)$-two-island network, where $V=V_1\cup V_2$, $|V_1|=|V_2|=n$, $|V|=2n$.
\begin{itemize}
    \item \textbf{Parameter settings}: $\forall i\in V$, $b_i=b> 0$, $\phi_i=\phi\in (0,1)$, and $m_{ij}=w_{ij}/d_i$ for $j\in V$.
    \item \textbf{Initial states}: $\forall i\in V_1$, $x_i(0)=x_0\in (\frac{1}{2},1)$ and $y_i(0)=y_0\in[\frac{1}{2}, x_0]$.
    $\forall j\in V_2$, $x_j(0)=1-x_0$ and $y_j(0)=1-y_0$. 
\end{itemize}
\end{assum}
\begin{rem}
Since two opposing groups of individuals participate in group discussion, to avoid being criticized or isolated for disagreeing with others, individuals express their opinions more euphemistically. Hence, we assume that the initial explicit opinions of
individuals are more neutral, i.e., $|y_i(0)-\frac{1}{2}|\leq |x_i(0)-\frac{1}{2}|$ for 
all $i\in V$. 
\end{rem}

Next, we will show the co-evolutionary process and results of dual opinions. Lemmas and proofs are in Appendices \ref{ap:lemma} an \ref{ap:th2}, respectively.

\begin{thm}\label{thm:x>y>0.5}
Consider a social network $G(W)$, where every individual's dual opinions evolve according to \eqref{eq:sys}. Suppose that Assumption \ref{ass} holds. Then, for individuals $i,j$ in the same group, $x_{i}(t)=x_{j}(t)$ and $y_{i}(t)=y_{j}(t)$ hold for any $t\geq 0$.
For individuals $i\in V_1$, $j\in V_2$, it holds that for any $t>0$, $x_i(t)+x_j(t)=y_i(t)+y_j(t)=1$, $x_i(t)> y_i(t)>0.5$, and $x_j(t)< y_j(t)<0.5$.
\end{thm}
The proof is simple and omitted here. Theorem \ref{thm:x>y>0.5} shows the evolutionary characteristic of dual opinions, and suggests that explicit opinions are always more neutral than their implicit opinions, which is also supported by empirical research \citep{stangor2016study}. 

We divide the bias parameter of individuals into three levels, precisely,
\vspace{-1cm}
\begin{small}
\begin{equation*}
    \begin{split}
        &\textbf{Extreme bias}: b\geq 1.\\
        &\textbf{Moderate bias}: \frac{2}{\phi(h_G-1)+2}< b<1.\\
        &\textbf{Slight bias}: 0<b\leq\frac{2}{\phi(h_G-1)+2}.
    \end{split}
\end{equation*}
\end{small}
Next, we will introduce the co-evolution of dual opinions with extreme, moderate, and slight bias respectively. 
{Concisely, we will only describe the results of group $V_1$, and the results of group $V_2$ can be obtained according to Theorem \ref{thm:x>y>0.5} similarly.}

%%%------ b\geq 1 -----------------------
\subsection{Extreme bias}
\begin{thm}\label{thm:extreme}
{Consider a social network $G(W)$, where individuals have extreme bias, i.e., $b\geq 1$, and their dual opinions evolve as \eqref{eq:sys}.
Suppose that Assumption \ref{ass} holds.
Then, for individual $i\in V_1$, $x_i(t+1)> x_i(t)$ holds for any $t \geq 0$, and there exits $\tau>0$ such that $y_i(t)$ is monotonic when $t> \tau$. Additionally, } 
\vspace{-0.5cm}
\begin{small}
\begin{equation*}
    \lim_{t\to\infty}x_i(t)=1, \text{ and } \lim_{t\to\infty}y_i(t)=\frac{\phi h_G+1}{\phi h_G+2-\phi}, \forall i\in V_1.
\end{equation*}
\end{small}

\end{thm}
\vspace{-0.2cm}

The proof is inspired by \cite{dandekar2013biased} and
shown in Appendix \ref{ap:th2}.
Theorem \ref{thm:extreme} indicates that 
when individuals are extremely biased, even under great social pressure, the implicit opinion formation is polarizing and the implicit opinion eventually converge to the extreme states, despite individuals expressing a relatively neutral explicit opinion.   Fig.\ref{fig:y0}(d) shows the process of dual opinions formation when individuals are extremely biased. As shown in Fig.\ref{fig:y0}(d), group discussion will deepen the differences in true thoughts between two opposing groups,
which is also found in empirical research \citep{myers1970discussion}.

Besides, if individuals are extremely biased, the limits of explicit opinions satisfy $\lim_{t\to\infty}y_l(t)\neq \lim_{t\to\infty}x_l(t)$ and $\lim_{t\to\infty}y_l(t)\neq\frac{1}{2}$ for every individual $l\in V$, which reveals that explicit opinions will not converge to the extreme states nor achieve the consensus, and explicit opinions of individuals in opposing groups will remain in persistent disagreement.

\subsection{Moderate bias}
Define a function
\begin{scriptsize}
\begin{equation*}
    g(x,b)=\left\{
    \begin{aligned}
        ~&\frac{x^{1-b}-(1-x)^{1-b}}{x(1-x)^{1-b}-(1-x)x^{1-b}},& x\in(\frac{1}{2},1], 0<b<1, \\
        ~&\frac{2}{b}-2, & x=\frac{1}{2}, 0<b<1. \\
    \end{aligned}\right.
\end{equation*}
\end{scriptsize}
Denote $\hat{x}(\phi, h_G, b)$ as the solutions of the equation
\vspace{-0.5cm}
\begin{small}
\begin{equation}\label{eq:equation g(x,b)}
    g(x,b) = \phi(h_G-1).
\end{equation}
\end{small}
When $\frac{2}{\phi(h_G-1)+2}< b<1$,  \eqref{eq:equation g(x,b)} {has a unique solution} $\hat{x}(\phi, h_G, b)\in (\frac{1}{2},1)$. The existence and uniqueness of $\hat{x}(\phi, h_G, b)$ are provided in Lemma \ref{lem:g(x,b)} (see Appendix \ref{ap:lemma}).

%%%%%%%%%%%%%% \frac{2}{\phi(h_G-1)+2}\leq b<1 %%%%%%%

\begin{thm}\label{thm:b<moderate}
Consider a social network, where individuals are with moderate bias, i.e., $\frac{2}{\phi(h_G-1)+2}< b<1$, and their dual opinions update according to \eqref{eq:sys}. 
Suppose that Assumption \ref{ass} holds. Then, for individual $i\in V_1$, there exists $\tau>0$ such that for any $t>\tau$, the dual opinions $x_i(t)$ and $y_i(t)$ are monotonic, and

\vspace{-0.6cm}
\begin{small}
\begin{align}
    \lim_{t\to\infty} x_i(t)&=\hat{x}(\phi, h_G, b),\label{eq:thmF:limitx:V}\\
    \lim_{t\to\infty} y_i(t)&=\frac{\phi(h_G+1)\hat{x}(\phi, h_G, b)+1-\phi}{\phi h_G+2-\phi}.\label{eq:thmF:limity:V1}
\end{align}
\end{small}
\end{thm}
\vspace{-0.2cm}

The proof is presented in Appendix \ref{ap:th2}. { Theorems \ref{thm:extreme} and \ref{thm:b<moderate} } suggest that if individuals have a { extreme or moderate} bias, not only individuals in the social network will not reach a consensus, but their own implicit and explicit opinions will also remain in persistent disagreement.  Due to social pressure, individuals express explicit opinions that { conform more} to public opinion, showing the phenomenon of outward conformity called compliance, as in \cite{myers_social_2010}. 

\cite{cheng2021social,luo2022Adapted} combine the BC model with the EPO model framework and also shows the phenomenon of compliance. Since the opinion update rules in \cite{cheng2021social,luo2022Adapted} are based on the weighted average algorithm, as the group discussion progresses, the biggest and smallest opinions will become more neutral, 
which fails to capture the opinion polarization.
In the dual opinion co-evolution model, when individuals are extremely biased, group discussion will strengthen the initial preference of individuals, which means that implicit opinions will become more extreme and eventually converge to the extreme states, and the gap between the largest and smallest opinions will increase as opinions evolve.

\subsection{Slight bias}
%%%------- \frac{2}{h_G+1}\leq b< \frac{2}{\phi(h_G-1)+2} ------

\begin{thm}\label{thm:slight}
Consider a social network $G(W)$, where individuals have slight bias, i.e., $b\leq \frac{2}{\phi(h_G-1)+2}$, and their dual opinions evolve according to \eqref{eq:sys}.
Suppose that Assumption \ref{ass} holds. Then, for individual $i\in V_1$, there exists $\tau>0$ such that for any $t>\tau$, $x_i(t+1)\leq x_i(t)$ and $y_i(t+1)\leq y_i(t)$ hold. Additionally, 
\vspace{-0.3cm}
\begin{small}
\begin{equation*}
    \lim_{t\to\infty}x_i(t)=\frac{1}{2}, \text{ and } \lim_{t\to\infty}y_i(t)=\frac{1}{2}, \forall i\in V_1.
\end{equation*}
\end{small}
Especially, if $b<\frac{2}{h_G+1}$, $\tau$ can be $0$.
\end{thm}

The proof is presented in Appendix \ref{ap:th2}.
Theorem \ref{thm:slight} suggests that when individuals are slightly biased, implicit and explicit opinions converge to the public opinion $\frac{1}{2}$,  
showing a dual opinions formation process where individuals gradually accept and agree with the public opinion through the group discussion.

Besides, when $\frac{2}{h_G+1}\leq b\leq \frac{2}{\phi(h_G-1)+2}$, the implicit and explicit opinions may not be monotone at the beginning, but as the group discussion progresses, the dual opinions monotonically converge to $\frac{1}{2}$. 
As the bias parameter decreases,
i.e., when $b<\frac{2}{h_G+1}$,  the dual opinions monotonically converge to the neutral state $0.5$.

To summarize, Theorems \ref{thm:extreme}-\ref{thm:slight} show that $b=1$ and $b=\frac{2}{\phi(h_G-1)+2}$ are two critical values that divide the degree of individual bias into three levels.  When the bias parameter increases from slight bias to moderate or even extreme bias, individuals will {change} from sincere acceptance to outward compliance. When individuals are extremely biased, the implicit opinion formation process is polarizing though individuals publicly conform to the public opinion.

\begin{rem}\label{rmk:two_con}
In the standard biased assimilation model \citep{dandekar2013biased}, i.e., without explicit opinions, opinions evolve monotonically, so the existence of limits of opinions can be directly deduced by the monotone bounded convergence theorem.
The co-evolution of dual opinions is more complex. Implicit and explicit opinions are coupled and may not be monotonic at first, as shown in Fig. \ref{fig:y0}(c). Even more, as shown in Fig. \ref{fig:y0}(d), the implicit opinion value is increasing while the explicit opinion value is decreasing.
Therefore, the proof in \cite{dandekar2013biased} can not be used directly in our derivation.
\end{rem}
\vspace{-0.15cm}
\subsection{The influence of parameters on the co-evolution of dual opinions}

Let $\mathcal{P}=(y_0,\phi^\mathcal{P}, b^\mathcal{P}, h_G^\mathcal{P})$ present the parameter settings for the model \eqref{eq:sys} under Assumption \ref{ass}. Denote $x_i^\mathcal{P}(t)$ and $y_i^\mathcal{P}(t)$ as the implicit and explicit opinions of individual $i\in V$ at time $t$ in the group discussion, where parameters in the update rules \eqref{eq:sys} are given in $\mathcal{P}$. 

\begin{thm}\label{thm:parameter}
Consider two group discussions on a social network $G(W)$ with parameter settings $\mathcal{P}_1$ and $\mathcal{P}_2$, respectively, and dual opinions evolve by \eqref{eq:sys}. Suppose that Assumption \ref{ass} holds. If for any individuals $l\in V$, $x_l^{\mathcal{P}_1}(0)=x_l^{\mathcal{P}_2}(0)$ and $\mathcal{P}_1>\mathcal{P}_2$ hold, then
for any $t>0$,
\vspace{-0.3cm}
\begin{small}
\begin{equation}\label{eq:thm5:1}
    x_i^{\mathcal{P}_1}(t)\geq x_i^{\mathcal{P}_2}(t) \text{ and } y_i^{\mathcal{P}_1}(t)\geq y_i^{\mathcal{P}_2}(t), \forall i\in V_1.
\end{equation}
\end{small}
\end{thm}
\vspace{-0.2cm}
The proof is shown in Appendix \ref{ap:th2}.  

Theorem \ref{thm:parameter} shows how individual resilience $\phi$, the biased parameter $b$, the degree of homophily $h_G$, and the initial explicit opinion $y_0$ influence the evolutionary process of implicit and explicit opinions {under Assumption \ref{ass}}. Specifically, the initial explicit opinion only influences the dual opinions formation process and does not influence the limits of dual opinions, and the increased social pressure $1-\phi$ prompts dual opinions to be more neutral.

When individuals have moderate bias, the limits of implicit and explicit opinions depend on parameters $h_G$, $\phi$ and $b$.  Lemma \ref{lem:g(x,b)} (see Appendix \ref{ap:lemma}) shows that {under Assumption \ref{ass}}, the limits of dual opinions will become more extreme if the parameters $b$, $h_G$, or $\phi$ increase. When individuals are extremely biased, the limits of explicit opinions depend on the degree of homophily of the two-island network $h_G$ and social pressure $1-\phi$. If the limits of explicit opinions of individuals in a high-homophily two-island network are expected to be the same as in a low-homogeneity network, greater social pressure from the social network needs to be placed on individuals.

\section{Numerical simulations}\label{sec:numerical}

This section will present some numerical simulations to verify our theoretical results and show two types of conformity, namely acceptance and compliance. 

\begin{figure}
    \centering
    \includegraphics[width=8cm]{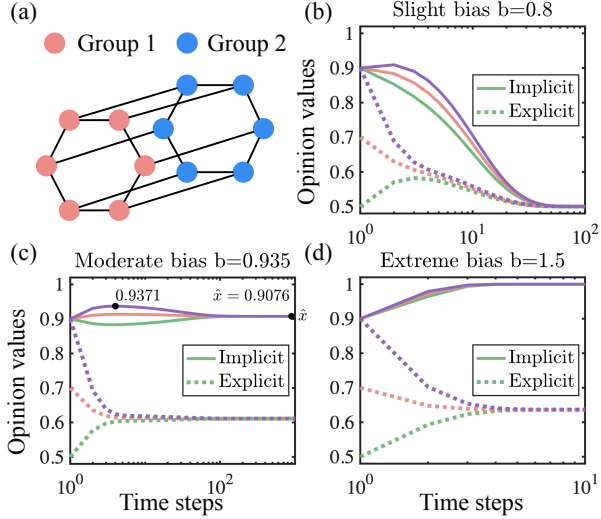}
    \caption{(a) The $(6,6,2/6,1/6)$-two-island network $G(W)$. (b)-(d) The trajectories of implicit and explicit opinions 
     of agent in group $V_1$. The solid lines represent implicit opinions while dashed lines represent explicit opinions. Initial explicit opinions $y_0=0.9, 0.7, 0.5$ are colored in purple, pink, green.
    }
    
    \label{fig:y0}
\end{figure}     
% fig1117.eps

%\
Consider two opposing groups of individuals discussing a topic on the $(6,6,2/6,1/6)$-two-island network $G(W)$, shown in Fig \ref{fig:y0}(a), where Assumption \ref{ass} holds and $\phi=0.2$.
Hence, the bias parameter $b=\frac{10}{11}$ % 2/2.2
and $b=1$ are two critical values to grade the degree of individual bias.
Figs. \ref{fig:y0}(b)-(d) show the trajectories of dual opinions of individuals in group $V_1$.

In Fig.\ref{fig:y0}(b) where individuals are slightly biased, the dual opinions reach a consensus at the neutral state $0.5$, showing the phenomenon of acceptance, where individuals sincerely agree with the public opinion. 
Figs. \ref{fig:y0}(c)-(d) show that when individuals have moderate or extreme bias, individuals will act in compliance with public opinion, i.e.,  individuals express explicit opinions more in line with public opinion while
they do not accept it inside. When individuals are extremely biased, as shown in Fig. \ref{fig:y0}(d), the group discussion strengthens individuals' initial preference resulting in polarization, i.e, the implicit opinions of individuals in group $V_1$ are polarizing to the extreme state $1$.

Figs.\ref{fig:y0}(b)-(d) show that the initial explicit opinion $y_0$ affects the evolution process of dual opinions, that is, the bigger the initial explicit opinion $y_0$, the bigger the implicit and explicit opinions in the evolution process, which verify the validity of Theorem \ref{thm:parameter}. Besides, the implicit opinions and explicit opinions may even evolve in opposing directions. As shown by the purple line $(y_0=0.9)$ in Fig. \ref{fig:y0}(d), the implicit opinion is increasing while the explicit opinion is decreasing.

Implicit and explicit opinions may not be monotone initially but eventually monotonically converge. 
Specifically, in Fig.\ref{fig:y0}(c), implicit opinions of individuals in group $V_1$ converge to 
$\hat{x}(0.2, 2, 0.935)=0.9076$ (keep 4 valid digits). 
Looking at the evolution trajectory (the solid purple line) of the implicit opinion, we can see that the implicit opinion increases at the beginning, even beyond the limit $\hat{x}(0.2, 2, 0.935)$, and then monotonically decreases, and ultimately converges.

%\

%%%%%%%%%%%%%%%%%%%% Conclusion %%%%%%%%%%%%%%
\section{Conclusion}\label{sec:conclusion}

This paper proposes an opinion dynamics model to investigate
the co-evolution of dual opinions.  The implicit and explicit opinions co-exist on an object. 
Theoretical analysis 
has exhibited the co-evolution of dual opinions on a two-island network in an adversarial discussion scenario, where two groups of individuals with opposite initial opinions
participate in the discussion. As the bias parameter increases, individuals will change from acceptance to compliance. Further, we have rigorously derived the critical values of transition between the above patterns of evolutionary results.
In addition, we have analyzed the impact of parameters on the co-evolution process and results under {the assumption}.
 
In this paper, implicit opinions and explicit opinions are updated synchronously. According to the dual attitudes model theory, implicit opinions change more slowly compared with explicit opinions. To capture the phenomenon more accurately, we plan to modify the model by separating update time scales of implicit and explicit opinions and conduct further research.

\vspace{-0.3cm}

%diversified environment
\begin{ack}                               % Place acknowledgements
This work was supported by the National Natural Science Foundation of China (No. 61873167), and the Project of Science and Technology Commission of Shanghai Municipality, China under Grant 22JC1401401, and in part by Shanghai Talent Development Fund and  Shuguang Scholar of Shanghai.
\end{ack}

       % Include this if you use bibtex 
\bibliography{autosam}

%%%%%%%%%%%%%%%%%%%% Appendices %%%%%%%%%%%%%
\appendix

%%%%%%%%%%%% Appendix lemma %%%%%%%%%%%%%%%%%

\section{Notations of Appendices}\label{app:organization}

Theorem \ref{thm:x>y>0.5} shows that individuals in the same group have the same dual opinions and individuals in different groups have the opposite dual opinions. Hence, we will only prove the results of individual $i\in V_1$ and the results of individuals in group $V_2$ can be proved in a similarly way by using Theorem \ref{thm:x>y>0.5}.

{For clarity, we abbreviate ``for every individual $i\in V_1$ (or $V_2$, $V$)" as ``for $i\in V_1$ (or $V_2$, $V$)".} 
Denote
\vspace{-0.5cm}
\begin{small}
\begin{equation*}
\begin{split}
    & f(u)=\frac{u}{1-u}, u\in [0.5,1)\\
    &S(x,\phi)=n(p_s-p_d)\frac{\phi(h_G+1)x+1-\phi}{\phi h_G+2-\phi}+np_d,\\
    & z_i(t)=\frac{x_i(t)}{1-x_i(t)}, d_i=np_s+np_d:=d,  \forall i\in V.
\end{split}
\end{equation*}
\end{small}
In addition to the proof of Lemma \ref{lem:g(x,b)}, denote $\hat{x}(\phi)$ as $\hat{x}(\phi, h_G, b)$.
\section{Lemmas and corresponding proofs}\label{ap:lemma}

\begin{lem}\label{lem:facts}
Suppose that Assumption \ref{ass} holds and dual opinions evolve as \eqref{eq:sys}.
For individual $i\in V_1$ and $t\geq 0$, the following facts hold:
\begin{enumerate}[F1.]
    \item\label{f1} $z_i(t+1)=z_i(t)^bf(\frac{s_i(t)}{d_i}) \geq1$;
    \item\label{f2} ${1\leq}f(\frac{s_i(t)}{d})=\frac{h_Gy_i(t)+1-y_i(t)}{y_i(t)+h_G-h_Gy_i(t)}{\leq \frac{h_Gx_i(t)+1-x_i(t)}{x_i(t) + h_G-h_Gx_i(t)}}$;
    \item\label{f3} if $y_i(t+1)<y_i(t)$, it holds that $s_i(t+1)<s_i(t)$ and $f(\frac{s_i(t+1)}{d})<f(\frac{s_i(t)}{d})$;
    \item\label{f4} if $y_i(t+1)\geq y_i(t)$, it holds that 
\begin{scriptsize}
\begin{equation}\label{eq:ap_s&S}
    f(\frac{s_i(t+1)}{d})\leq f(\frac{S(x_i(t+1),\phi)}{d});
\end{equation}
\end{scriptsize}
\vspace{-0.2cm}
{
    \item\label{f5} if $g(x_i(t),b)< \phi(h_G-1)$ (or $>$), it holds that 
    \begin{scriptsize}
    \begin{equation}\label{eq:ap:f5}
        z_i(t)< z_i(t)^bf(\frac{S(x_i(t+1),\phi)}{d})~ (\text{or} >);
    \end{equation}
    \end{scriptsize}
    \item\label{f6} if $g(x_i(t),b)\geq (h_G-1)$, it holds that 
    \begin{scriptsize}
    \begin{equation}\label{eq:ap:f6}
        z_i(t)\geq z_i(t)^b\frac{(h_G-1)x_i(t)+1}{(h_G-1)(1-x_i(t))+1}.
    \end{equation}
    \end{scriptsize}} 
\end{enumerate}  
\end{lem}
\vspace{-0.5cm}

{
\textbf{Proof of Lemma \ref{lem:facts}.}
According to the update rules \eqref{eq:sys} and Theorem \ref{thm:x>y>0.5}, it's easily to prove Facts F\ref{f1}-F\ref{f3}. For $i\in V_1$, $\hat{y}_{i, avg}(t)=\frac{1}{h_G+1}(h_Gy_i(t)+1-y_i(t))$, which follows that $y_i(t+1)=\phi x_i(t+1) + \frac{(1-\phi)}{h_G+1}(h_Gy_i(t)+1-y_i(t))\geq y_i(t)$. Consequently, we obtain \eqref{eq:ap_s&S}. Rearranging the inequality $g(x_i(t),b)< \phi(h_G-1)$ yields that $x_i(t)^{1-b}(\phi(h_G-1)(1-x_i(t))+1)< (1-x_i(t))^{1-b}(\phi(h_G-1)x_i(t)+1)$, thus we can get $z_i(t)^{1-b}< \frac{\phi(h_G-1)x_i(t)+1}{\phi(h_G-1)(1-x_i(t))+1}=f(\frac{S(x_i(t+1),\phi)}{d})$, i.e., \eqref{eq:ap:f5} holds. Taking $\phi=1$, we have \eqref{eq:ap:f6}.\qed
}

\begin{lem}\label{lem:g(x,b)}
Given that $x\in[\frac{1}{2},1]$, $b\in (0, 1)$, $\phi\in (0,1)$ and $h_G>1$. Then, the following statements hold.\\
$(\romannumeral1)$ The function $g(x,b)$ is continuous.\\
$(\romannumeral2)$ The function $g(x,b)$ is increasing with respect to variable $x$ and is decreasing with respect to variable $b$.\\
$(\romannumeral3)$ If $\frac{2}{\phi(h_G-1)+2}< b<1$, $\hat{x}(\phi, h_G, b)\in (\frac{1}{2},1)$ is the unique solution of equation $g(x,b)=\phi(h_G-1)$ and is increasing with respect to variables $b, h_G, \phi$.
\end{lem}

\textbf{Proof of Lemma \ref{lem:g(x,b)}.}
First, we will show that $g(x,b)$ is a continuous function when $x\in [\frac{1}{2},1]$. We only need to check the continuity of $g(x,b)$ at $x=\frac{1}{2}$.
By the L'Hospital's rule, we get that $\lim_{x\to\frac{1}{2}}g(x,b)=\frac{2}{b}-2=g(\frac{1}{2},b)$. Hence, $g(x,b)$ is continuous at $x=\frac{1}{2}$. 

Second, we will show that $g(x,b)$ is increasing with respect to the variable $x$. 
Let $\frac{g_1(x,b)}{g_2(x,b)}=\frac{\partial g(x,b)}{\partial x}$, where $g_1(x,b)=(1-b)x^{2-b}(1-x)^{-b}-x^{2-2b}+(1-x)^{2-2b}-(1-b)x^{-b}(1-x)^{2-b}$, and $g_2(x,b)=\big{[}x(1-x)^{1-b}-(1-x)x^{1-b}\big{]}^2$. Then, we have
$\frac{\partial g_1(x,b)}{\partial x}\geq 0$ when $x\in (\frac{1}{2},1]$. Thus, $g_1(x,b)$ is increasing with respect to the variable $x$. It follows that $\frac{\partial g(x,b)}{\partial x}=\frac{g_1(x,b)}{g_2(x,b)}\geq \frac{g_1(\frac{1}{2},b)}{g_2(x,b)}=0$.
Accordingly, $g(x,b)$ is increasing with respect to $x$. 
In addition, $\frac{\partial g(x, b)}{\partial b}\leq 0$, which follows that $g(x,b)$ is decreasing with respect to the variable $b$. 

Third, assume that $\frac{2}{\phi(h_G-1)+2}< b<1$. Then, we get $g(\frac{1}{2},b)=\frac{2}{b}-2< \phi(h_G-1)<g(1,b)=+\infty$.
By the intermediate value theorem, there exists a unique solution $\hat{x}(\phi, h_G, b)\in (\frac{1}{2},1)$ such that $g(\hat{x}(\phi, h_G, b),b)=\phi(h_G-1)$. 
Assume that $b_1<b_2$. Statement ($\romannumeral2$) shows that $g(x,b_1)\leq g(x, b_2)$, and then we get that $\hat{x}(\phi, h_G, b_1)\geq \hat{x}(\phi, h_G, b_1)$, which reveals that $\hat{x}(\phi, h_G, b)$ is decreasing with respect to $b$. Besides, as $\phi$ or $h_G$ increases, the value of $\phi(h_G-1)$ increases, and thus $\hat{x}(\phi, h_G, b)$ increases. \qed

\begin{rem}
    If $\frac{2}{\phi(h_G-1)+2}<b<1$, Lemma \ref{lem:g(x,b)} shows that $g(\hat{x}(\phi), b)=\phi(h_G-1)$. {Rearranging the above equation, we have}
\vspace{-0.2cm}
\begin{scriptsize}
\begin{equation}\label{eq:ap_r(p)}
    f(\hat{x}(\phi))=f(\hat{x}(\phi))^bf(\frac{S(\hat{x}(\phi),\phi)}{d}).
\end{equation}
\end{scriptsize}
\end{rem}

\begin{lem}\label{lem:xy_limit}
Given that $b\in (0,1)$, $\phi\in(0,1)$ and $h_G> 1$.
Suppose that $(x,y)$, $\frac{1}{2}\leq y\leq x\leq 1$, is the solution of the following nonlinear equations
\vspace{-0.3cm}
\begin{scriptsize}
\begin{equation}\label{eq:xysolutions}
    \left\{
    \begin{aligned}
    &x=\frac{x^b(h_G y+1-y)}{x^b(h_G y+1-y)+(1-x)^b(y+h_G-h_G y)} \\
    &y=\phi x+\frac{1-\phi}{h_G+1}(h_G y+1-y)
    \end{aligned},
    \right.
\end{equation}
\end{scriptsize}
Then, if $0<b\leq \frac{2}{\phi(h_G-1)+2}$, the solution set of \eqref{eq:xysolutions} is $\mathcal{S}=\{(\frac{1}{2},\frac{1}{2}), (1, \frac{\phi h_G+1}{\phi h_G+2-\phi})\}$; otherwise, the solution set is $\mathcal{S}\cup \{(\hat{x}(\phi), \frac{\phi(h_G+1)\hat{x}(\phi)+1-\phi}{\phi h_G+2-\phi})\}$.
\end{lem}

\textbf{Proof of Lemma \ref{lem:xy_limit}.}
It's easy to verify that $(\frac{1}{2},\frac{1}{2})$ and $(1, \frac{\phi h_G+1}{\phi h_G+2-\phi})$ are solutions of \eqref{eq:xysolutions}. So, it remains to show that if $b> \frac{2}{\phi(h_G-1)+2}$, there remains only one solution $(\hat{x}(\phi), \frac{\phi(h_G+1)\hat{x}(\phi)+1-\phi}{\phi h_G+2-\phi})$, and if $b\leq \frac{2}{\phi(h_G-1)+2}$, \eqref{eq:xysolutions} have no other solutions. Assume that $x\in (\frac{1}{2},1)$. Rearranging the equations \eqref{eq:xysolutions} yields $g(x,b)=\phi(h_G-1)$ and $y=\phi x+\frac{1-\phi}{h_G+1}(h_G y+1-y)$.
If $b> \frac{2}{\phi(h_G-1)+2}$, Lemma \ref{lem:g(x,b)} shows that there exists a unique solution $\hat{x}(\phi)$ such that $g(\hat{x}(\phi),b)=\phi (h_G-1)$. It implies that $(\hat{x}(\phi), \frac{\phi(h_G+1)\hat{x}(\phi)+1-\phi}{\phi h_G+2-\phi})$ is the only solution of the equations \eqref{eq:xysolutions} when 
$x\in (\frac{1}{2},1)$. If $b\leq \frac{2}{\phi(h_G-1)+2}$, then $g(\frac{1}{2},b)=\frac{2}{b}-2\geq\phi(h_G-1)$. Because $g(x,b)$ is  increasing when $x\in(\frac{1}{2}, 1)$, $g(x,b)>g(\frac{1}{2},b)\geq \phi(h_G-1)$ holds. Therefore, if $0<b\leq \frac{2}{\phi(h_G-1)+2}$, \eqref{eq:xysolutions} has no extra solutions except for $(\frac{1}{2}, \frac{1}{2})$ and $(1, \frac{\phi h_G+1}{\phi h_G+2-\phi})$. \qed

\begin{lem}\label{lem:y_monotonic}
Suppose that Assumption \ref{ass} holds. For {individual} $i\in V_1$, if there exists $k_x$ such that $x_i(t)$ is monotonic for all $t>k_x$, then there exists $k>k_x$ such that $y_i(t)$ is monotonic for all $t>k$, and $x_i(t)$, $y_i(t)$ converge.
\end{lem}

\textbf{Proof of Lemma \ref{lem:y_monotonic}.} 
Without loss of generality, assume that for individual $i\in V_1$, $x_i(t+1)\geq x_i(t)$ holds for all $t> k_x$. We will prove by contradiction that there exists $k>k_x$ such that for individual $i\in V_1$, $y_i(t)$ is monotonic for all $t>k$. Assume that there exists a time sequence $\{t_l\}_{l=0}^{\infty}>k_x$ such that for individual $i\in V_1$, $y_i(t_{2l}+1)>y_i(t_{2l})$ and $y_i(t_{2l+1}+1)<y_i(t_{2l+1})$ hold. 
For individual $i\in V_1$, $y_i(t_0+1)>y_i(t_0)$ implies that $\hat{y}_{i, avg}(t_0+1)>\hat{y}_{i, avg}(t_0)$, and consequently we get $y_i(t_0+2)-y_i(t_0+1)>\phi(x_i(t_0+2)-x_i(t_0+1))\geq 0$. By iteration, we can prove that for $i\in V_1$, $y_i(t+1)>y_i(t)$ holds for any $t>t_0$, which contradicts with the assumption. Therefore, there exists $k>k_x$ 
such that for individual $i\in V_1$, $y_i(t)$ is monotonic for $t> k$. By Theorem \ref{thm:x>y>0.5}, we obtain that 
for individual $l\in V$ and $t>k$, 
$x_l(t)$ and $y_l(t)$ is monotonic, and thus $x_l(t)$ and $y_l(t)$ converge.  \qed

\begin{lem}\label{lem:x_geq_hat_x_phi}
Suppose that $\frac{2}{\phi(h_G-1)+2}< b<1$, and Assumption \ref{ass} holds. 
For individual $i\in V_1$, 
if there exists $T>0$ such that $\hat{x}(\phi)\leq x_i(t)< \hat{x}(1)$ holds for $t\geq T$, then there exists $\mathcal{T}>T$ such that $x_i(t)$ and $y_i(t)$ are monotonic for $t\geq \mathcal{T}$. Additionally, \eqref{eq:thmF:limitx:V} and \eqref{eq:thmF:limity:V1} hold.
\end{lem}

\textbf{Proof of Lemma \ref{lem:x_geq_hat_x_phi}.} First, we will show by contradiction that there exists $T_x>T$ such that for individual $i\in V_1$, $x_i(t)$ is monotonic for $t\geq T_x$.
Assume that there exists a time sequence $\{k_l\}_{l=0}^{\infty}>T$ such that 
for individual $i\in V_1$, $x_i(k_{2l}+1)<x_i(k_{2l})$ and $x_i(k_{2l+1}+1)>x_i(k_{2l+1})$ hold. 
The proof is complete if we can prove that 
no such $\{k_l\}_{l=0}^{\infty}$ exists. Along this way, we will show by induction that $x_i(t+1)\leq x_i(t)$ holds for individual $i\in V_1$. Base case: $x_i(k_{0}+1)<x_i(k_{0})$ holds for individual $i\in V_1$.
Inductive step: Assume that for individual $i\in V_1$, $x_i(k+1)\leq x_i(k)$ holds for some $k>k_0$, i.e., $z_i(k+1)\leq z_i(k)$.
If $y_i(k+1)<y_i(k)$ holds for $i\in V_1$, then $f(\frac{s_i(k+1)}{d})< f(\frac{s_i(k)}{d})$ holds by {Fact F\ref{f3}}, and thus $z_i(k+2)=(z_i(k+1))^bf(\frac{s_i(k+1)}{d})< (z_i(k))^bf(\frac{s_i(k)}{d})=z_i(k+1)$ holds, i.e., $x_i(k+2)<x_i(k+1)$.
Otherwise, if $y_i(k+1)\geq y_i(k)$ holds for $i\in V_1$, by {Fact F\ref{f4}}, one gets $f(\frac{s_i(k+1)}{d})\leq f(\frac{S(x_i(k+1),\phi)}{d})$, and thus {$z_i(k+2)=z_i(k+1)^bf(\frac{s_i(k+1)}{d})\leq z_i(k+1)^bf(S(x_i(k+1),\phi))\leq z_i(k+1)$, where the last inequality holds by Fact F\ref{f5} since $g(x_i(k+1),b)\geq g(\hat{x}(\phi),b)=\phi(h_G-1)$ holds.} 
To summary, $x_i(k+2)\leq x_i(k+1)$ holds for individual $i\in V_1$, which completes the inductive proof. Therefore, for individual $i\in V_1$ and $t>k_0$, $x_i(t+1)\leq x_i(t)$ holds, which contradicts with the assumption. Hence, there exists $T_x>T$ such that for individual $i\in V_1$, $x_i(t)$ is monotonic for $t\geq T_x$. 

Lemma \ref{lem:y_monotonic} shows that there exists $T_y>T_x$ such that for individual $i\in V$, $y_i(t)$ is monotonic for $t>T_y$, and $x_i(t)$, $y_i(t)$ converge. For individual $i\in V_1$, since $\hat{x}(\phi)\leq x_i(t) <\hat{x}(1)$ holds, Lemma \ref{lem:xy_limit} suggests that $\lim_{t\to\infty}x_i(t)=\hat{x}(\phi)$, thus \eqref{eq:thmF:limitx:V}-\eqref{eq:thmF:limity:V1} hold. \qed

%----Lemma 6------------
\begin{lem}\label{lem:x_leq_hat_x_phi}
Suppose that $\frac{2}{\phi(h_G-1)+2}\leq b<1$, and Assumption \ref{ass} holds. 
For {individual} $i\in V_1$, if there exists $T>0$ exists such that for $t\geq T$, $x_i(t)< \hat{x}(\phi)$ holds, then there exists $\mathcal{T}>T$ such that $x_i(t)$ and $y_i(t)$ are monotonic for $t\geq \mathcal{T}$. Additionally, \eqref{eq:thmF:limitx:V} and \eqref{eq:thmF:limity:V1} hold.
\end{lem}

\textbf{Proof of Lemma \ref{lem:x_leq_hat_x_phi}.} 
Similar to the proof of Lemma \ref{lem:x_geq_hat_x_phi}, we can prove by contradiction that there exists $\tau_x>T$ such that for $i\in V_1$, $x_i(t)$ is monotonic for $t\geq \tau_x$.

Then, we will show by contradiction that for individual $i\in V_1$, $x_i(t+1)\geq x_i(t)$ holds for all $t\geq \tau_x$. Assume that for individual $i\in V_1$, $x_i(t+1)< x_i(t)$ holds for all $t\geq \tau_x$. 
If there exists $k>\tau_x$ such that $y_i(k+1)\leq y_i(k)$ {holds for individual $i\in V_1$}, then we can obtain  
$f(\frac{s_i(k+1)}{d})\geq f(\frac{S(x_i(k+1),\phi)}{d})$ by {Fact F\ref{f4}}, thus {$z_i(k+2)=z_i(k+1)^bf(\frac{s_i(k+1)}{d})\geq z_i(k+1)^bf(\frac{S(x_i(k+1),\phi)}{d})=z_i(k+1)^b\frac{\phi(h_G-1)x_i(k+1)+1}{\phi(h_G-1)(1-x_i(k+1))+1}> z_i(k+1)$}, {where the last inequality holds by Fact F\ref{f5} since $g(x_i(k+1),b)> g(\hat{x}(\phi),b)=\phi(h_G-1)$ holds.} Consequently, 
$x_i(k+2)> x_i(k+1)$ holds for individual $i\in V_1$ leading to a contradiction. Otherwise, i.e., {for individual $i\in V_1$, $y_i(t+1)>y_i(t)$ holds for all $t>\tau_x$, we obtain that $y_i(t)$ is monotonic for all $t>\tau_x$, thus  $y_i(t)$ converges. Denote $x^*=\lim_{t\to\infty}x_i(t)$ and $y^*=\lim_{t\to\infty}y_i(t), i\in V_1$.} Thus, the limits $x^*$ and $y^*$ should satisfy: $x^*<\hat{x}(\phi)$ and $y^*>\frac{1}{2}$. However, Lemma \ref{lem:xy_limit} suggests that no such solution exists that satisfies the above constraints, which leads to a contradiction. To sum up, our assumption must be wrong. Hence, for individual $i\in V_1$ and $t\geq \tau_x$, $x_i(t+1)\geq x_i(t)$. Based on this,
Lemma \ref{lem:y_monotonic} suggests that for individual $i\in V_1$, $x_i(t)$ and $y_i(t)$ converge. Then, by Lemma \ref{lem:xy_limit}, we have $\lim_{t\to\infty}x_i(t)=\hat{x}(\phi)$ for individual $i\in V_1$, thus \eqref{eq:thmF:limitx:V}-\eqref{eq:thmF:limity:V1} hold. \qed

\section{Proofs of Theorems \ref{thm:extreme}-\ref{thm:parameter}} \label{ap:th2}

\textbf{Proof of Theorem \ref{thm:extreme}.} 
For individual $i\in V_1$, we have $z_i(t+1)=z_i(t)^bf(\frac{s_1(t)}{d})>z_i(t)$ for $t>0$ {by using Facts F\ref{f1}-\ref{f2} in Lemma \ref{lem:facts}}, which follows that $x_i(t+1)>x_i(t)$, {i.e., $x_i(t)$ is increasing for $t>0$.} 
According to Theorem \ref{thm:x>y>0.5} and Lemma \ref{lem:y_monotonic}, there exists $\tau> 0$ such that for individual $i\in V_1$, $y_i(t)$ is monotonic for $t>\tau$, and $x_i(t)$, $y_i(t)$ converge. Let $x^*$ and $y^*$ represent the limits of $x_i(t)$ and $y_i(t)$ for individual $i\in V_1$ so that $x^*\geq x_i(t)>\frac{1}{2}$ holds for all $t>0$, which follows that $x_i(t)(1-x_i(t))\geq x^*(1-x^*)$. Let $C(x)=x^{b-1}-(1-x)^{b-1}$, where $x\in [\frac{1}{2}, 1]$. One gets that $\frac{\partial C(x)}{\partial x}>0$, which implies that for $i\in V_1$ and $t>0$, $C(x^*)\geq C(x_i(t))>C(x_0)$. 

Next, we will show by contradiction that $x^*=1$. Assume that $x^*<1$. For $\epsilon=\frac{x^*(1-x^*)C(x_0)}{2}$, since for $i\in V_1$, $x_i(t)$ converges to $x^*$, there exists $T_\epsilon>0$ such that for $t>T_\epsilon$, $|x_i(t)-x^*|<\epsilon$. It follows that
\begin{equation} \label{eq:x_i(t+1)-x_i(t)<2epsilon}
    |x_1(t+1)-x_1(t)|<2\epsilon.
\end{equation}
Then, we get that for individual $i\in V_1$,
\vspace{-0.5cm}
\begin{small}
\begin{equation*}\label{eq:x(t+1)-x(t)>2e}
    \begin{split}
        &x_i(t+1)-x_i(t)\\
        =&\frac{x_i(t)\big{(}1-x_i(t)\big{)}\Big{[}x_i(t)^{b-1}s_i(t)-\big{(}1-x_i(t)\big{)}^{b-1}\big{(}d-s_i(t)\big{)}\Big{]}}{x_i (t)^b s_i(t)+\big{(}1-x_i(t)\big{)}^b\big{(}d-s_i(t)\big{)}}\\
        >&x_i(t)\big{(}1-x_i(t)\big{)}\Big{[}x_i(t)^{b-1}-\big{(}1-x_i(t)\big{)}^{b-1}\Big{]}\\
        >&x^*(1-x^*)C(x_0)=2\epsilon,
    \end{split}
\end{equation*}
\end{small}
which contradicts with assumption \eqref{eq:x_i(t+1)-x_i(t)<2epsilon}.
The first inequality holds since {$x_i (t)^b s_i(t)+\big{(}1-x_i(t)\big{)}^b\big{(}d-s_i(t)\big{)}<s_i(t)$} and $s_i(t)>d-s_i(t)$ hold for $i\in V_1$ and $t\geq 0$. Consequently, for individual $i\in V_1$, $x^*=1$ holds.
Taking the limit of both sides of \eqref{eq:sys}, we have $y^*=\frac{\phi h_G+1}{\phi h_G+2-\phi}$. Now, the proof of Theorem \ref{thm:extreme} is completed.\qed\\

\vspace{-0.5cm}

\textbf{Proof of Theorem \ref{thm:b<moderate}.} Before proceeding with the proof, we need to prove three claims first.

\begin{claim}\label{clm:1}
    For individual $i\in V_1$, if there exists $k>0$ such that $x_i(k)<\hat{x}(1)$ holds, then $x_i(t)<\hat{x}(1)$ for all $t\geq k$.
\end{claim}

For individual $i\in V_1$, if $x_i(k)< \hat{x}(1)$ holds, then {$z_i(k+1)=z_i(k)^bf(\frac{s_i(k)}{d})<f(\hat{x}(1))^bf(\frac{S(\hat{x}(1),1)}{d})=f(\hat{x}(1))$, where the last inequality holds by \eqref{eq:ap_r(p)}}, thus $x_i(k+1)< \hat{x}(1)$ holds. An easy induction gives that for individual $i\in V_1$, $x_i(t)< \hat{x}(1)$ holds for all  $t\geq k$. \qed

\begin{claim}\label{clm:2}
    For individual $i\in V_1$, if there exists $k\geq 0$ such that $x_i(k)\geq \hat{x}(1)>\hat{x}(\phi)$ holds, then there exists $T>k$ such that $x_i(t)< \hat{x}(1)$ holds for all $t> T$.
\end{claim}

By Claim \ref{clm:1}, for individual $i\in V_1$ if there exists $T>0$ such that $x_i(T)${$<$}$\hat{x}(1)$, then $x_i(${$t)<$}$\hat{x}(1)$ holds for $t>T$. So, we only need to prove that for individual $i\in V_1$, there exists finite $T>k$ such that $x_i(T)<\hat{x}(1)$ holds, which will be proved by contradiction.
Assume that for individual $i\in V_1$, $x_i(t)\geq \hat{x}(1)$ holds for all $t\geq k$. 
Then, for $i\in V_1$ and $t>k$, we have
{ 
$z_i(t+1)=z_i(t)^bf(\frac{s_i(t)}{d}) 
< z_i(t)^b\frac{h_Gx_i(t)+1-x_i(t)}{x_i(t) + h_G-h_Gx_i(t)} 
\leq z_i(t)$,
where the first inequality holds by Fact F\ref{f2}, and the second inequality holds by Fact F\ref{f6} since $g(x_i(t),b)\geq g(\hat{x}(1),b)=h_G-1$ holds. By iteration, we can obtain that for $i\in V_1$ and $t>k$, $x_i(t+1)<x_i(t)$ holds, i.e., $x_i(t)$ is decreasing when $t>k$}. Lemma \ref{lem:y_monotonic} suggests that for individual $i\in V_1$, $x_i(t)$ and $y_i(t)$ converge. Let $x^*$ represent the limit of $x_i(t), i\in V_1$ so that $x^*$ should satisfy $\hat{x}(1)\leq x^* <1$.
However, Lemma \ref{lem:xy_limit} shows that no such solution exists meeting the above condition. Therefore, our assumption must be wrong, which reveals that 
there exists finite $T>k$ such that for individual $i\in V_1$, $x_i(T)<\hat{x}(1)$ holds and by Claim 1, $x_i(t)< \hat{x}(1)$ holds for all $t>T$. \qed

\begin{claim}\label{clm:3}
    For individual $i\in V_1$, if there exists $k\geq 0$ such that $\hat{x}(\phi)\leq x_i(k)< \hat{x}(1)$ holds, then only one of the following situations will happen: \\
S1. $\forall t> k$, $\hat{x}(\phi)\leq x_i(t)< \hat{x}(1)$;\\
S2. $\exists T_x\geq k$ such that $\forall t\geq T_x$, $x_i(t)< \hat{x}(\phi)$.
\end{claim}

We will show it by contradiction. Assume that there exists a time sequence $\{k_l\}_{l=0}^{\infty}\geq k$ such that for $i\in V_1$, $x_i(k_{2l}+1)<\hat{x}(\phi)< x_i(k_{2l})$ and $x_i(k_{2l+1})<\hat{x}(\phi)< x_i(k_{2l+1}+1)$ hold. Contradiction appears if no such time sequence $\{k_l\}_{l=0}^{\infty}$ exists. The above assumption suggests that $z_i(k_0+1)<f(\hat{x}(\phi))<z_i(k_0)$ holds for all $i\in V_1$.  
 
\textbf{C1.} If $y_i(k_0+1)<y_i(k_0)$ for $i\in V_1$, one gets that $z_i(k_0+2)=z_i(k_0+1)^bf(\frac{s_i(k_0+1)}{d})<z_i(k_0)^bf(\frac{s_i(k_0)}{d})=z_i(k_0+1)$, {where the first inequality holds by Fact F\ref{f3}},
thus we have $x_i(k_0+2)< x_i(k_0+1)< \hat{x}(\phi)$. {Based on this, for $i\in V_1$, $y_i(k_0+2)-y_i(k_0+1)<(1-\phi)(\hat{y}_{i,avg}(k_0+1)-\hat{y}_{i,avg}(k_0))<0$ holds, where the last inequality holds since $y_i(k_0+1)<y_i(k_0)$.}
An easy induction gives that for $i\in V_1$, $x_i(t+1)<x_i(t)<\hat{x}(\phi)$ and $y_i(t+1)<y_i(t)$ hold for any $t>k_0$, which leads to a contradiction.

\textbf{C2.} Otherwise, i.e., $y_i(k_0+1)\geq y_i(k_0), i\in V_1$, we have $f(\frac{s_i(k_0+1)}{d})\leq f(\frac{S(x_i(k_0+1),\phi)}{d})< f(\frac{S(\hat{x}(\phi),\phi)}{d})$, {where the first inequality holds by Fact F\ref{f4}, and the second by the assumption that $x_i(k_0+1)<\hat{x}(\phi)$}, 
thus $z_i(k_0+2)< f(\hat{x}(\phi))^bf(\frac{S(\hat{x}(\phi),\phi)}{d})=f(\hat{x}(\phi))$, {where the equality holds by \eqref{eq:ap_r(p)}.}
i.e., $x_i(k_0+2)< \hat{x}(\phi)$ holds. For individual $i\in V_1$,
\begin{enumerate}
    \item[C21.]if $x_i(k_0+2)\leq x_i(k_0+1)$, $y_i(k_0+2)<y_i(k_0+1)$, then similar to C1, the proof is complete;
    \item[C22.]if $x_i(k_0+2)\leq x_i(k_0+1)$, $y_i(k_0+2)\geq y_i(k_0+1)$, then similar to C2, we obtain that $x_i(t_0+3)< \hat{x}(\phi)$; 
    \item[C23.]if $x_i(k_0+2)> x_i(k_0+1)$, it must be $y_i(k_0+1)\geq y_i(k_0)$, which implies $y_i(k_0+2)\geq y_i(k_0+1)$. Similar to C2,  $f(s_i(k_0+2))\geq f(s_i(k_0+1))$ $\Longrightarrow$  $x_i(k_0+3)<\hat{x}(\phi)$ and $x_i(k_0+3)>x_i(k_0+2)$.
\end{enumerate}
Iterative calculation above yields that for individual $i\in V_1$, $x_i(t)<\hat{x}(\phi)$ for all $t>k_0$, which leads to a contradiction. Therefore, our assumption must be wrong, which completes the proof of Claim 3. \qed

Now, we start the main proof of Theorem \ref{thm:b<moderate}, which will be divided into two cases.

\textbf{Case 1.} $x_i(0)\geq \hat{x}(\phi)$ for $i\in V_1$.
Claims \ref{clm:1}-\ref{clm:2} imply that there exists $k\geq 0$ such that for individual $i\in V_1$, $x_i(k)< \hat{x}(1)$ holds for any $t>k$.  Claim 3 suggests that only one of situations S1 and S2 holds, and the proofs are shown in  Lemmas \ref{lem:x_geq_hat_x_phi} and \ref{lem:x_leq_hat_x_phi} respectively.

\textbf{Case 2.} $x_i(0)<\hat{x}(\phi)$ for $i\in V_1$. If there exists $k >0$ such that for individual $i\in V_1$,  $x_i(k)\geq \hat{x}(\phi)$, then the theorem can be proved by the analysis in Case 1.
Otherwise, i.e., for individual $i\in V_1$, $x_i(t)<\hat{x}(\phi)$ holds for all $t\geq 0$, the proof is shown in Lemma \ref{lem:x_leq_hat_x_phi}.

Now the proof of Theorem \ref{thm:b<moderate} ends.\qed\\

\vspace{-0.5cm}

\textbf{Proof of Theorem \ref{thm:slight}.} The proof will be divided into the following three cases: 

\textbf{Case 1.} $\frac{2}{h_G+1}\leq b \leq \frac{2}{\phi(h_G-1)+2}$, $x_i(0)\leq \hat{x}(1)$, $\forall i\in V_1$.

\textbf{Step c1.1} We will show by induction that for individual $i\in V_1$, $x_i(t)\leq \hat{x}(1)$ holds for all $t\geq 0$. Base case: $x_i(0)\leq \hat{x}(1)$ holds for $i\in V_1$. Induction step: assume that for individual $i\in V_1$, $x_i(k)\leq \hat{x}(1)$ holds. {It follows that for $i\in V_1$, $ f(\hat{x}(1))=f(\hat{x}(1))^bf(\frac{S(\hat{x}(1),1)}{d})> z_i(k)^bf(\frac{s_i(k)}{d})=z_i(k+1)$ holds, where the first equality holds by \eqref{eq:ap_r(p)} and $>$ holds since $\hat{x}(1)\geq x_i(k)>y_i(k)$.} Hence, $x_i(k+1)<\hat{x}(1)$ for $i\in V_1$, which complete the inductive proof.

\textbf{Step c1.2} We will show that there exists $\tau_x\geq 0$ such that for  $i\in V_1$, $x_i(t)\leq x_i(\tau_x)$ holds for all $t>\tau_x$. 

First, we will show by contradiction that there exists $\tau_x\geq 0$ such that for individual $i\in V_1$, $x_i(\tau_x+1)\leq x_i(\tau_x)$ holds. Assume that for individual $i\in V_1$, $x_i(t+1)>x_i(t)$ holds for all $t\geq 0$. Lemma \ref{lem:y_monotonic} implies that for $i\in V_1$, $x_i(t)$ and $y_i(t)$ converge. Denote $x^*=\lim_{t\to\infty}x_i(t)$ for $i\in V_1$ so that  $x^*>\frac{1}{2}$ since $x_i(t+1)>x_i(t)$ for all $t\geq 0$. 
However, Lemma \ref{lem:xy_limit} shows that no solution exists fulfilling the above condition. Therefore, our assumption must be wrong, that is, there exists $\tau_x\geq 0$ such that $x_i(\tau_x+1)\leq x_i(\tau_x)$ holds for individual $i\in V_1$.

Second, we will show by induction that for individual $i\in V_1$, $x_i(t+1)\leq x_i(t)$ holds for any $t\geq\tau_x$. Base case: $x_i(\tau_x+1)\leq x_i(\tau_x)$ holds for $i\in V_1$. Induction step: assume that for individual $i\in V_1$, 
$x_i(k+1)\leq x_i(k)$ holds, where $k\geq \tau_x$. For individual $i\in V_1$, if $y_i(k+1)< y_i(k)$, similar to C1 in Claim \ref{clm:3}, we can obtain that  $x_i(t+1)< x_i(t)$ holds for $t>k$; otherwise, i.e., $y_i(k+1)\geq y_i(k)$, we have {$z_i(k+2)=z_i(k+1)^bf(\frac{s_i(t+1)}{d})\leq z_i(k+1)^bf(\frac{S(x_i(k+1),\phi)}{d})<z_i(k+1)$, where the first inequality holds by Fact F\ref{f4}, and the second inequality holds by Fact F\ref{f5} since $g(x_i(k+1),b)> g(\frac{1}{2},b)=\frac{2}{b}-2>\phi(h_G-1)$ holds.} Thus, for $i\in V_1$, $x_i(t+2)\leq x_i(t+1)$ holds for all $t>\tau_x$, which completes the inductive proof. 

\textbf{Step c1.3} We will prove that for $i\in V_1$, $x_i(t)$ and $y_i(t)$ converge to $\frac{1}{2}$.
Since for individual $i\in V_1$, $x_i(t+1)\leq x_i(t)$ holds for any $t\geq \tau_x$, Lemma \ref{lem:y_monotonic} suggests that there exists $\tau_y>\tau_x$ such that $y_i(t)$ is monotonic for all $t>\tau_y$, and $x_i(t)$ and $y_i(t)$ converge. Let $x^*$ and $y^*$ represent the limits of $x_i(t)$ and $y_i(t)$  for $i\in V_1$ so that $x^*=y^*= \frac{1}{2}$ holds by Lemma \ref{lem:xy_limit},  which in turn reveals that for $t\geq \tau_y$, $y_i(t+1)\leq y_i(t)$. 
Let $\tau=\max\{\tau_x,\tau_y\}$. For individual $i\in V_1$, $x_i(t+1)\leq x_i(t)$ and $y_i(t+1)\leq y_i(t)$ holds for all $t\geq \tau$, which completes the proof of Case 1.

\textbf{Case 2.} $\frac{2}{h_G+1}\leq b<\frac{2}{\phi(h_G-1)+2}$, $x_i(0)>\hat{x}(1)$, $\forall i\in V_1$.

First, we will show by contradiction that there exists $T>0$ such that $x_i(T)\leq \hat{x}(1)$ holds.
Assume that for $i\in V_1$, $x_i(t)>\hat{x}(1)$ holds for all $t\geq 0$. 
It follows that for $i\in V_1$ and $t\geq 0$, {$z_i(t)>z_i(t)^b\frac{h_Gx_i(t)+1-x_i(t) }{h_G-h_Gx_i(t)+x_i(t)}>z_i(t)^bf(\frac{s_i(t)}{d})=z_i(t+1)$, where the first inequality holds by Fact F\ref{f6} since $g(x_i(t),b)> g(\hat{x}(1),b)=h_G-1$ holds, and the second holds since $y_i(t)<x_i(t)$ by Theorem \ref{thm:x>y>0.5}, i.e.,
$x_i(t)>x_i(t+1)>\hat{x}(1)$.} 
Lemma \ref{lem:y_monotonic} shows that for individual $i\in V_1$, $x_i(t)$ and $y_i(t)$ converge so that $\lim_{t\to\infty}x_i(t) \geq \hat{x}(1)$. However, by Lemma \ref{lem:xy_limit}, no such solution exists satisfying the above condition, which leads to a contradiction. Therefore, our assumption must be wrong. Thus, there exists $T>0$ such that $x_i(T)\leq \hat{x}(1)$ holds for individual $i\in V_1$. Then, the proof is complete by the proof of Case 1.

\textbf{Case 3.} $b<\frac{2}{h_G+1}$.

\vspace{-0.2cm}
For $i\in V_1$ and $t\geq 0$,  {
$z_i(t)>z_i(t)^b \frac{h_G x_i(t)+1-x_i(t)}{x_i(t)+h_G(1-x_i(t))}\geq z_i(t)^bf(\frac{s_i(t)}{d})=z_i(t+1)$ holds, where the first inequality holds by Fact F\ref{f6} since $g(x_i(t),b)> g(\frac{1}{2},b)=\frac{2}{b}-2\geq h_G-1$ holds, and the second inequality holds since $x_i(t)\geq y_i(t)$ by Theorem \ref{thm:x>y>0.5}.}
It implies that for individual $i\in V_1$,  $x_i(t+1)<x_i(t)$ holds for any $t>0$. Lemma \ref{lem:y_monotonic} shows that there exists $\tau>0$ such that for $i\in V_1$, $y_i(t)$ is monotonic for $t>\tau$, and $x_i(t)$, $y_i(t)$ converge. By Lemma \ref{lem:g(x,b)}, for individual $i\in V_1$,  $\lim_{t\to\infty}x_i(t)=\frac{1}{2}$ and $\lim_{t\to\infty}y_i(t)=\frac{1}{2}$ hold.
Hence, the proof of Theorem \ref{thm:slight} is complete. \qed

\textbf{Proof of Theorem \ref{thm:parameter}.} Consider two discussions with parameter settings $\mathcal{P}_1=(y^{\mathcal{P}_1}_0,\phi^{\mathcal{P}_1}, b^{\mathcal{P}_1}, h_G^{\mathcal{P}_1})>\mathcal{P}_2=(y^{\mathcal{P}_2}_0,\phi^{\mathcal{P}_2}, b^{\mathcal{P}_2}, h_G^{\mathcal{P}_2})$.
For individual $i\in V_1$, denote $s_i^{\mathcal{P}}(t)=\sum_{j\in N_i}w_{ij}y_j^{\mathcal{P}}(t)$, and $\hat{y}^{\mathcal{P}}_{i, avg}=\sum_{j\in N_i}m_{ij}y_j^{\mathcal{P}}(t)$, and $z_i^{\mathcal{P}}(t)=\frac{x_i^{\mathcal{P}}(t)}{1-x_i^{\mathcal{P}}(t)}$, and then we get that
$s_i^{\mathcal{P}_1}(0)\geq s_i^{\mathcal{P}_2}(0)$ and  $\hat{y}_{i, avg}^{\mathcal{P}_1}(0)\geq \hat{y}_{i, avg}^{\mathcal{P}_2}(0)$ and $b^{\mathcal{P}_1}\geq b^{\mathcal{P}_2}$ and $\phi^{\mathcal{P}_1}\geq \phi^{\mathcal{P}_2}$ hold, and at least one of the above inequalities holds the greater-than sign $>$. It follows that inequalities \eqref{eq:thm5:1} hold at $t=1$. 
By iteration, one gets that \eqref{eq:thm5:1} holds for all $t>0$, which completes the proof. \qed

%---------------------------------------------------

\end{document}